\documentclass[11pt,a4paper,twocolumn,floatfix]{quantumarticle}
\pdfoutput=1

\usepackage[utf8]{inputenc}
\usepackage[english]{babel}
\usepackage[T1]{fontenc}
\usepackage{amsmath}
\usepackage{amssymb}
\usepackage{amsfonts}
\usepackage{bm}
\usepackage{graphicx}
\usepackage{xcolor}
\usepackage{tikz}
\usepackage{algpseudocode}
\usepackage{tabularx}
\usepackage{booktabs}
\usepackage{adjustbox}
\usepackage{multirow}
\usepackage[numbers,sort&compress]{natbib}
\usepackage{hyperref}

\newtheorem{proposition}{Proposition}
\newtheorem{corollary}{Corollary}

\newcounter{algocounter}
\renewcommand{\thealgocounter}{\arabic{algocounter}}
\renewcommand{\algref}[1]{Alg.~\ref{algo:#1}}

\newcommand{\citeasnoun}[1]{Ref.~\cite{#1}}
\newcommand{\figref}[1]{Fig.~\ref{fig:#1}}

\newcommand{\figrefbegin}[1]{Figure~\ref{fig:#1}}
\newcommand{\tableref}[1]{Table~\ref{tab:#1}}

\newcommand{\appref}[1]{Appendix~\ref{app:#1}}
\renewcommand{\eqref}[1]{Eq.~(\ref{eq:#1})}

\begin{document}

\title{Degree-Guided Adiabatic Quantum Algorithm for MIS with Neutral Atoms}

\author{Guy Karni}
\affiliation{Racah Institute of Physics, Hebrew University of Jerusalem, Jerusalem 9190401, Israel}

\author{Noam Cohen}
\affiliation{School of Computer Science and Engineering, Hebrew University of Jerusalem}

\author{Adi Pick}
\email{adi.pick@mail.huji.ac.il}
\affiliation{Institute of Applied Physics, Hebrew University of Jerusalem, Jerusalem 9190401, Israel}

\maketitle

\begin{abstract}
We present a quantum adiabatic algorithm for solving MIS with Rydberg atoms, which uses graph degrees to engineer  the Hamiltonian for adiabatic evolution. Our approach extends the algorithm of Pichler et al. for MIS state preparation~\cite{pichler2018quantum}, which uses  time-dependent global controls to excite atoms while the Rydberg blockade mechanism suppresses adjacent excitations. To accelerate adiabatic evolution, we introduce local light shifts that favor excitation of low-degree nodes and constrain the final Hamiltonian so that energy scales inversely with the number of Rydberg excitations, preventing convergence to non-MIS states. Benchmarking our  degree-dependent MIS algorithm with the traditional adiabatic algorithm on a set of 1000 MIS instances, we demonstrate numerically  that our algorithm inflates  the minimal spectral gap and, consequently,  accelerates the  convergence.  By fitting power-law relations for both the time-to-solution and success probability metrics as a function of the hardness parameter, we demonstrate a superior scaling exponent for our algorithm over the traditional global protocol. Furthermore, the classical Hamiltonian engineering step required to construct degree-dependent controls introduces only polynomial overhead.  These  results indicate  that our   algorithm can be used to  extend the size of graphs for which MIS can be found on physical hardware. Moreover, our approach extends beyond MIS to additional graph problems such as MAX-CUT.
\end{abstract}

\section{Introduction}\label{sec:introduction}

Neutral atom arrays are a promising platform for solving graph optimization problems~\cite{ebadi2022quantum,graham2022multi,kim2022rydberg,nguyen2023quantum}. In \citeasnoun{pichler2018quantum}, Pichler introduced an  adiabatic quantum algorithm (AQC) for solving the  maximum independent set (MIS) problem, which seeks to find the largest subset of mutually non-adjacent vertices in a graph.   The Pichler algorithm and related  adiabatic-inspired algorithms  were implemented on Rydberg arrays containing hundreds of atoms~\cite{bernien2017probing,ebadi2022quantum,kim2024quantum,de2025demonstration}. To scale up the performance  for larger systems, a challenge arises since adiabatic evolution requires controls that vary slowly relative to the energy gap~\cite{farhi2000quantum,albash2018adiabatic}, but the latter  typically shrinks upon increasing the graph size~\cite{sachdev1999quantum}. \emph{In this paper, we propose a modification of the Pichler algorithm that uses an efficient Hamiltonian engineering step prior to running the adiabatic control sequence and achieves faster convergence to the MIS solution.}

\begin{figure}[t]
    \centering
    \includegraphics[width=\columnwidth]{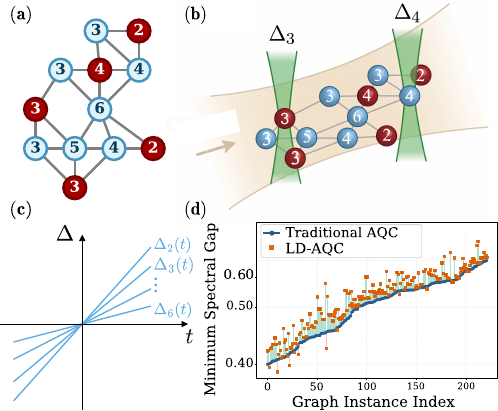}
    \caption{\textbf{LD-AQC MIS algorithm with degree-dependent detunings: } (a) Example  graph with MIS  colored in red and degree labels on vertices. (b) Rydberg qubit array with in-plane global control beams (orange shading) and localized perpendicular beams (green shading) that induce degree-dependent light shifts required for the LD-AQC. For clarity, perpendicular beams are shown on only two sites. (c) Degree-dependent detunings as a function of time. (d) Minimum spectral gap of a set of Hamiltonians [\eqref{LD_hamiltonian}] representing 1000 MIS problems, using traditional AQC with global drives (blue) and LD-AQC with  degree-dependent drives  (orange).}
    \label{fig:figure1}
\end{figure}

Our approach relies  on the observation that lower-degree vertices are more likely to belong to the MIS, as selecting  a low-degree vertex as a MIS member excludes fewer adjacent nodes and helps maximize the set size.  However, degree cannot be  the sole  criterion for MIS selection. As shown in \figref{figure1}(a), vertices with identical degrees can be either inside (red) or outside (blue) the MIS.  Thus, naively selecting all nodes of minimum degree   yields a wrong MIS  solution.  To resolve this challenge, we systematically tailor the frequency profiles of the local driving pulses, preferentially exciting low-degree atoms while avoiding erroneous state preparation [see~\figref{figure1}(b, c)].

To explain how we achieve this,  let us describe the adiabatic MIS algorithm. As in the Pichler approach, we excite the atoms via rapid adiabatic passage (RAP)~\cite{allen2012optical}, sweeping the pulse detuning from negative to positive while the Rydberg blockade prevents neighboring excitations. However, in contrast to global drives used in the Pichler algorithm, we engineer degree-dependent detunings that favor the excitation of low-degree vertices~\figref{figure1}(b-c). To prevent erroneous excitations, we require that these local detunings preserve an energy ordering in the final-stage Hamiltonian [\eqref{energy-cond}], ensuring that states with more excitations maintain lower energies throughout the evolution. We show that this construction  increases the minimum spectral gap, thereby accelerating the convergence~\figref{figure1}(d).

Before presenting our algorithm --- hereafter denoted the LD-AQC --- we briefly review the development of existing MIS AQC algorithms. While the original Pichler work from ~\citeasnoun{pichler2018quantum} suggested adiabatic following of ground states~\cite{farhi2000quantum}, the experimental realization in \citeasnoun{ebadi2022quantum} used  gate-based adiabatic-inspired  algorithms, including quantum approximate optimization algorithm (QAOA)~\cite{farhi2014quantum} and variational quantum adiabatic algorithm (VQAA)~\cite{schiffer2022adiabatic}, to analyze planar graphs with hundreds of vertices. These results were extended to non-planar graphs and various lattice geometries~\cite{dalyac2022embedding,dalyac2023exploring,andrist2023hardness,kim2024quantum,pan2025encoding}. Using polynomial mappings of MIS to other problems in NP, adiabatic algorithms were developed for the weighted MIS problem (WMIS)~\cite{de2025demonstration}, integer factorization~\cite{nguyen2023quantum}, and energy optimization in electric charging~\cite{leclerc2025implementing}. Since the time required for adiabatic evolution generally increases with the system's  size, various methods  were proposed for accelerating adiabatic protocols. These include optimizing the adiabatic schedule numerically~\cite{koch2022quantum,turyansky2024inertial} or analytically~\cite{roland2002quantum,turyansky2026pulse}, and the introduction of counter-diabatic controls~\cite{berry2009transitionless,sels2017minimizing,guery2019shortcuts}. Quantum algorithms that accelerate AQC using heuristic initial states are particularly related to the present work~\cite{crosson2021prospects}. In~\citeasnoun{perdomo2011study}, an algorithm for k-SAT introduced heuristic initial states along with increased fluctuations that enable escape from local minima while approaching the solution. A similar approach was used to prepare the ground state of the $p$-spin model, known as ``reversed quantum annealing''~\cite{developers2021reverse,chancellor2017modernizing}. A related approach of randomized initial states in AQC was explored in~\cite{chancellor2017modernizing,king2019quantum} and realized experimentally with the D-Wave processor~\cite{developers2021reverse}. Finally, we mention classical MIS algorithms that utilize vertex degrees. A widely used greedy algorithm constructs the MIS by iteratively selecting minimum-degree vertices, removing them along with their neighbors, and repeating on the reduced graph~\cite{halldorsson1994greed}. Another example is a branch-and-reduce method, which constructs the MIS recursively using a degree-dependent graph branching rule~\cite{fomin2006measure}.  \emph{Motivated by the speedup gained by heuristic AQC algorithms and that of degree-based classical MIS algorithms, we develop a degree-based heuristic for accelerating the Pichler  MIS algorithm.}

\section{The AQC MIS algorithm}\label{sec:hamiltonian}
 
Let us introduce the  rotating-frame Hamiltonian of the Rydberg qubit array with driving  pulses:
\begin{equation}
    H = \sum_{i \in V} \left[\Omega(t) \hat{\sigma}_x^{(i)} - \Delta_{i}(t) \hat{n}_i\right] + \sum_{(i,j) \in E} U_{ij} \hat{n}_i \hat{n}_j.
    \label{eq:LD_hamiltonian}
\end{equation}
The first sum corresponds to RAP pulses, while the second accounts for Rydberg interactions. The operators $\hat\sigma_x^{(i)}$ and $\hat{n}_i$ are the Pauli matrix and projector onto the Rydberg state of the $i^{\mathrm{th}}$ atom. The Rabi frequency $\Omega(t)$ vanishes at the start and end of the protocol, while the local detuning of the $i^\text{th}$ atom, $\Delta_i(t)$, changes from negative to positive along the evolution.  The second term represents Van-der Waals  interactions between atom pairs, with  $U_{ij}=C_6/r_{ij}^6$, where $r_{ij}$ is the distance between atoms~\cite{saffman2010quantum}. While the original AQC from \citeasnoun{pichler2018quantum} used global drives with homogeneous detunings, $\Delta_i(t)=\Delta_0(t)\,\forall i$, we construct degree-dependent local detunings. Implementation requires single-atom addressability, achievable with existing hardware~\cite{de2025demonstration,goswami2024solving}.

\section{Degree-dependent local detunings}\label{sec:ld-aqc}

In this section, we introduce an efficient algorithm for constructing degree-dependent  $\Delta_i(t)$, considering   functions with linear time dependence\footnote{ Although linear time dependence is  not optimal for RAP~\cite{roland2002quantum,albash2018adiabatic}, it is  convenient for analytic analysis.} [\figref{figure1}(c)].  Here we assume  that the detuning of vertex $i$ is of the form
\begin{equation}
    \Delta_i(t) = \delta_0 \cdot f(\beta,d_{i})  \left(\tfrac{t}{t_f} - \tfrac{1}{2}\right),
    \label{eq:local-detunings}
\end{equation}
where $\delta_0$ is an energy scale, $t_f$ is the protocol duration, and  $f(\beta,d_{i})$ is a monotonic function of the degree $d_i$, which depends on a parameter  $\beta$. 
We set  $f(\beta,d_j)=1$ for all vertices with $d_j=0$. 
Our goal is to find $\beta$ that accelerates the convergence towards the MIS solution.

\begin{figure}[h!]
    \centering
    \includegraphics[width=\columnwidth]{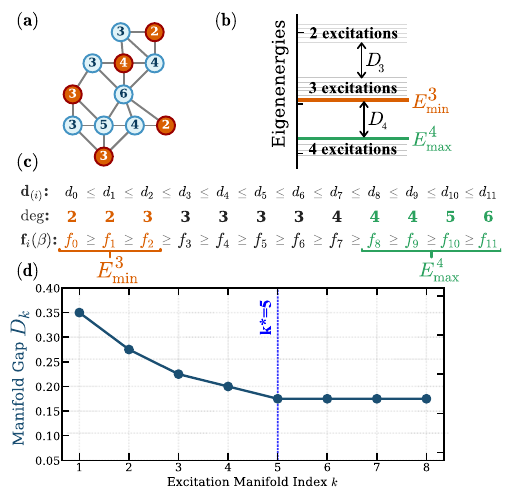}
    \caption{\textbf{Engineering degree-dependent detunings} (a) Example graph from \figref{figure1}(a). (b) Eigenenergies are grouped into bands corresponding to the  number of Rydberg excitations, and the energy gap  between excitation manifolds  $D_k\equiv E_\text{min}^{k-1} - E_\text{max}^{k}$ is shown.  (c) Vertices are sorted  by ascending degree order and corresponding  functions,  $f_k(\beta)$,  are assigned  in descending order.  (d) Energy gap function $D_k(\beta)$  is plotted for the linear function $f_k(\beta) = 1 - \beta\cdot d_k$. A critical value $k^*$ is shown, which is the minimal $k$ for which $D_{k^*}(\beta)\leq D_k(\beta)\,\forall\,k$. }
    \label{fig:figure2}
\end{figure}

A pseudocode of the algorithm for finding $\beta$ is shown in \algref{Algo1}. In steps 1--2, graph  vertices are sorted by ascending degree order and  the corresponding  detuning functions are sorted in descending order  [\figref{figure2}(c)].  Then,  the energy difference function, $D_k(\beta)$ is defined [\eqref{D_k_definition}], which is the energy gap between states with $k-1$ and $k$ excitations. Steps 3--5  find an optimal  $\beta^*$ value for which the energy difference function, $D_{k}(\beta^*)$,  is strictly non-negative for all $k$ values [satisfying \eqref{energy-cond}]. This condition implies that a  state with maximal number of Rydberg  excitations is the ground state. Therefore,  this step completes our Hamiltonian construction.

\begin{center}
\begin{minipage}{\columnwidth}
\refstepcounter{algocounter}\label{algo:Algo1}
\hrule\vspace{1.5mm}
\textbf{Algorithm \thealgocounter}\\
\emph{Given a parameter-dependent  function,  $f(\beta,d_i),$ the algorithm finds $\beta$ under which  the energy condition [\eqref{energy-cond}] is satisfied. Proofs  in \appref{proof-alg1}.}\vspace{0.15cm}
\hrule\vspace{1mm}
\begin{algorithmic}[1]
\State Sort vertices in ascending degree order, $\quad d_0\leq d_1\leq\dots d_{N-1}$.
\State Assign  detuning functions  in descending order, $f_0\geq f_1\geq\dots f_{N-1}$ 
\State Find minimal   $k^*$  so  that  $D_{k^*}(\beta)\leq D_k(\beta)\,\forall k$. 
\Statex \emph{Equivalently, find    minimal $k^*$ such that }
\Statex \emph{$d_{k^*-1} \ge d_{N-k^*-1}$ [proof in \eqref{energy-diff-equality}].}
\State Find $\beta_{c}$ such that $D_{k^*}(\beta_{c}) = 0$.
\State Choose $\beta^*<\beta_{c}$ to satisfy  $D_k(\beta^*) > 0, \forall\, k$.
\Statex \textbf{Output:} $\beta^*$
\end{algorithmic}
\vspace{1.5mm}\hrule
\end{minipage}
\end{center}

Next, we define $D_k(\beta)$. To this end, we observe that the  final-time Hamiltonian is approximately diagonal in the computational basis and takes the form  $H(t_f) \approx  - \tfrac{\delta_0}{2}\sum_{i} f(\beta,d_{i}) \hat{n}_i$\footnote{It follows since     the first and last terms in \eqref{LD_hamiltonian} vanish in the limit of infinite Rydberg blockade strength.}. Therefore,  the energy gap function is defined as 
\begin{gather}
    D_k(\beta) := \underbrace{\tfrac{\delta_0}{2}\sum_{i=N-k}^{N-1} f(\beta,d_{i})}_{E_{k}^\text{max}} - \underbrace{\tfrac{\delta_0}{2}\sum_{i=0}^{k-2} f(\beta,d_{i})}_{
     E_{k-1}^{\text{min}}},
    \label{eq:D_k_definition}
\end{gather}
where the first sum, containing the $k$ largest   values of $f(\beta,d_i)$, bounds the energy of states with $k$ excitations from above, while the second --- consisting of the $k-1$ smallest $f(\beta,d_i)$ values --- bounds the energy of states with $k-1$ excitations from below. Hence, $D_k(\beta)$ bounds the energy gap between the $k$ and $k-1$ excitation manifolds. The energy function, $D_k(\beta)$, for the specific choice of  $f_i(\beta) = 1 - \beta\cdot d_i$ is shown in \figref{figure2}(d). Additional nonlinear detuning functions are discussed in \appref{EM-DiffDetuning}.

Our algorithm finds an optimal parameter $\beta^*$ that satisfies \emph{the energy condition}:
\begin{equation}
D_k(\beta^*)>0 \quad \forall\, k = 1,\dots,N,
\label{eq:energy-cond}
\end{equation}
which implies that energies of states are sorted by the number of excitations.
Steps 3--5 in \algref{Algo1} are used  for  finding  $\beta^*$: Step 3 finds  the minimal $k^*$ for which   $D_{k^*}(\beta)\leq D_k(\beta) \,\forall k$. Then, step 4  finds a critical $\beta_{c}$ for which  $D_{k^*}(\beta_{c}) = 0$.  Finally, in step 5, choosing  $\beta^*<\beta_{c}$  implies that $D_{k}(\beta^*)> 0 \,\forall k$, as needed. In \appref{proof-alg1}, we discuss  propositions and assumptions  underlying \algref{Algo1}, including the existence of $k^*$ and $\beta^*$. 

By introducing degree-dependent detunings, we  lift degeneracies in the final-stage Hamiltonian. 
While with  global controls  ($\beta=0$), the eigenvalue spectrum of $H(t_f)$ consists of degenerate energy bands corresponding  to states with distinct numbers of Rydberg excitations, when $\beta\neq 0$, the degeneracy is lifted. Consequently, the  ground state of the non-degenerate Hamiltonian represents a solution to a  weighted MIS problem (WMIS)~\cite{de2025demonstration}. By imposing the energy condition, \eqref{energy-cond}, we require that the ground state of the WMIS problem coincides with that of the MIS, and that all energy bands remain gapped. 

Finally, we discuss  in \appref{blockade_check} a condition on the detuning sweep range set by the requirement that the blockade interaction strength must be sufficiently large relative to the final detunings to suppress neighboring excitations.

\emph{Complexity of Algorithm 1:} Given an input graph, $G(V,E)$ with $|V|=N$ vertices and $|E|=M$ edges, computing vertex degrees via an adjacency list takes $\mathcal{O}(N + M)$ operations, while sorting the vertices by degree requires $\mathcal{O}(N \log N)$ operations~\cite{cormen2022introduction}. Determining the critical $k^*$ requires a maximum of $N/2$ steps, while determining the critical $\beta^*$ requires solving $D_{k^*}(\beta^*) = 0$ via root-finding methods, e.g., Newton--Raphson~\cite{press2007numerical}. Summing the above contributions, the overall classical complexity of \algref{Algo1} is bounded by $\mathcal{O}(M + N \log N)$. Given that the maximum edge count is $M \le N(N-1)/2$, the complexity is bounded by $\mathcal{O}(N^2)$, implying a polynomial overhead.

\section{Performance analysis}\label{sec:benchmarking}

Having introduced  the algorithm for choosing local detunings in LD-AQC, we now benchmark its performance against traditional AQC using homogeneous detunings. To this end, we simulate both protocols to find approximate MIS solutions on King's graphs with random hole distributions.

We evaluate the average success probability under each algorithm,  defined as \begin{equation}
P_{\text{MIS}} \equiv 
\mathbb{E} \left[ \big\langle \psi(t_f) \big\vert   \hat{O}_{\text{MIS}}\big\vert \psi(t_f) \big\rangle \right],
\label{eq:P-mis-ensemble}
\end{equation}
where $\hat{O}_{\text{MIS}} = \sum_{k \in \mathcal{S}_{\text{MIS}}} \lvert \psi_k \rangle \langle \psi_k \rvert$ is the projection operator onto the MIS subspace,  $\mathbb{E}$ denotes an average over the graph ensemble, and $t_f$ is the final time.  Moreover, we compute  the approximation ratio error, defined as  $1-R$ with the approximation ratio defined as the ratio between the average excitation number and the MIS size:
\begin{equation}
R=\tfrac{1}{|\text{MIS}|}\mathbb{E}\!\left[\,\langle\psi(t_f)|\sum_{j\in G} \hat{n}_j|\psi(t_f)\rangle\right].
\label{eq:approximation-ratio}
\end{equation}

\figrefbegin{figure3}(a) shows the success probability  (blue) and approximation ratio error (red) achieved with the   traditional  and  LD-AQC algorithms on 1000 random graphs  with $11$ vertices.
Across all simulated protocol durations, the LD-AQC algorithm (dashed lines) consistently outperforms the traditional AQC (solid) in both metrics.  To simplify the simulations, we included Rydberg interactions only between nearest- and next-nearest neighbors, setting $U_{ij} = 0$ in \eqref{LD_hamiltonian} for all other atom pairs.

\begin{figure}[t]
    \centering
    \includegraphics[width=\columnwidth]{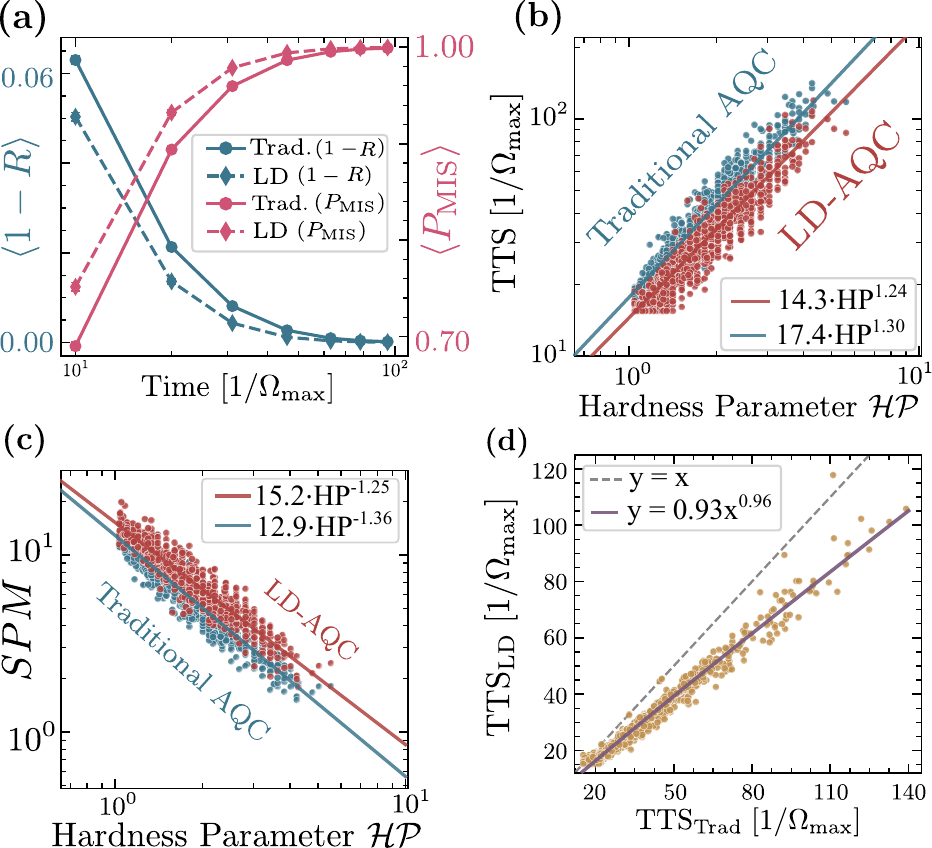}
    \caption{ \textbf{ Benchmarking performance of the traditional and LD-AQC algorithms. (a)} Average success probability, $P_{\text{MIS}}$ [\eqref{P-mis-ensemble}] (right axis), and approximation ratio error, $1- R$ [\eqref{approximation-ratio}] (left axis), versus  protocol duration $t_f$, averaged over 1000  MIS instances. \textbf{(b)} Scaling of the time-to-solution metric [\eqref{TTS} with  $p_{\mathrm{target}} = 0.99$] versus the traditional hardness parameter $\mathrm{HP}_{\mathrm{trad}}$ [\eqref{hardness_parameter_standard}], showing power-law fits for traditional AQC (blue) and LD-AQC (red). 
\textbf{(c)} Scaling of the success probability metric, $\mathrm{SPM}$ [\eqref{spm}], versus $\mathrm{HP}_{\mathrm{trad}}$ with power-law fits for both algorithms. 
\textbf{(d)} TTS  obtained with  LD-AQC versus that obtained with the traditional algorithm ($\mathrm{TTS}_{\mathrm{LD}}$ and  $\mathrm{TTS}_{\mathrm{trad}}$ respectively), plotted  on a log-log scale.}
    \label{fig:figure3}
\end{figure}

To analyze the complexity of our  algorithm, we studied  how  the performance metrics scale  with increasing  problem hardness.  Complexity of AQC algorithms is typically defined as the scaling of the time required to achieve a certain fidelity with problem size. A useful  measure of the time is the time-to-solution metric (or total annealing time), defined in~\citeasnoun{ronnow2014defining} as  
\begin{equation}
    \text{TTS}(t_f;p_\text{target}) = t_f \cdot \frac{ln(1- p_\text{target}) }{ln(1 - p_\text{MIS}(t_f))}, 
    \label{eq:TTS}
\end{equation}
given by the product of  a single-run duration time  and 
the number of repetitions needed to find the ground state at least once with
probability $p_\text{target}$, with $p_\text{MIS}(t_f)$ defined here per MIS instance (and not in the ensemble sense of \eqref{P-mis-ensemble}). The computational hardness of a MIS problem correlates not with graph size alone, but more accurately with the hardness parameter($\mathcal{HP}$) defined as
\begin{equation}
    \mathcal{HP}_{\text{trad}} = \frac{D_{\alpha-1}}{\alpha \cdot D_\alpha},
    \label{eq:hardness_parameter_standard} 
\end{equation}
where $\alpha(G)=|\text{MIS}|$ is the cardinality of the MIS, while $D_{j}$ is the number of independent sets  of size $j$ (i.e., the degeneracy). 

\figrefbegin{figure3}(b--c) show the scaling of the success probability metric, 
\begin{equation}
\text{SPM} = -\log(1-p_\text{MIS})
\label{eq:spm}
\end{equation}
and of the TTS with  $\mathcal{HP}_\text{trad}$ across one thousand random MIS instances. The TTS is computed using a required fidelity of $p_\text{target} = 0.99$. Both the TTS and the SPM scale linearly with $\mathcal{HP}_\text{trad}$ on a log-log scale, implying a power law relation between the quantities. Using a least-squares fit, we find that LD-AQC  achieves superior scaling compared to the traditional  AQC protocol  for the tested set of problems (with $R^2\approx0.9$ indicating good fit quality). As shown in \figref{figure3}(b), the LD-AQC algorithm consistently achieves lower TTS per MIS instance than the traditional AQC. Fitting to $\text{TTS} = C (\mathcal{HP}_\text{trad})^\gamma$ reveals that the traditional AQC has both a larger prefactor ($C$) and exponent ($\gamma$), indicating steeper scaling. Similarly, \figref{figure3}(c) demonstrates that LD-AQC yields a higher SPM that decays more slowly with increasing $\mathcal{HP}_\text{trad}$. Fitting to $\text{SPM} = C(\mathcal{HP}_\text{trad})^{-\gamma}$ confirms this trend, with LD-AQC exhibiting a larger $C$ and a smaller $\gamma$.
Finally, we use the same instances from panels (b,c) and plot  in \figref{figure3}(d) the  TTS of the LD-AQC versus that of the traditional algorithm. The least-mean-squares fit reveals a sublinear relation, with a power of $\text{TTS}_\mathrm{LD} = C(\text{TTS}_\text{trad})^{0.96}$, confirming the scaling advantage of our algorithm for this set of MIS instances (with $R^2\approx0.98$ indicating excellent  fit quality). Extending  this performance analysis to larger problem instances remains a direction for future work. We expect these favorable trends to persist, since the LD algorithm uses graph information to accelerate the convergence while at the same time  does not introduce significant bias towards erroneous solutions by maintaining the energy separation between manifolds [\eqref{energy-cond}].
In \appref{EM-PerformStas}, we show a histogram of the  logarithmic error ratio, $\log_{10} [ (1 - P_{\text{trad}}) / (1 - P_{\text{LD}}) ]$, for all MIS instances in the generated set of problems. We find that this ratio is positive for all instances in the set, showing that LD-AQC outperforms the traditional AQC on all instances.  This confirms that by satisfying the energy condition we avoid convergence to erroneous solutions.  Finally, we show in \appref{EM-DiffDetuning} that the LD protocol consistently outperforms traditional AQC for various detuning functions, $f(\beta,d_{i})$.

\section{Time complexity of the MIS  AQC}\label{sec:mechanism}

As mentioned in the previous section, the minimal gap in the traditional MIS AQC was shown to correlate with the hardness parameter, $\mathcal{HP}_\text{trad}$ [\eqref{hardness_parameter_standard}]. The latter depends on the degeneracy of states with $\alpha$ and $\alpha-1$ excitations. 
However, in the  LD-AQC algorithm,  the degeneracy of the energy bands is lifted. This calls for a modified definition of the HP. In this section, we construct a generalized HP [denoted $\mathcal{HP}_\text{LD}$ and defined in \eqref{hardness_parameter_LD}]. In  \appref{HP_anlyz_multy_vs_binary}, we show that across various correlation metrics, it exhibits stronger correlations with the minimal gap than the traditional HP. Therefore, it can be used to analyze the time complexity of the LD-AQC algorithm.

\begin{figure}[t]
    \centering
    \includegraphics[width=\columnwidth]{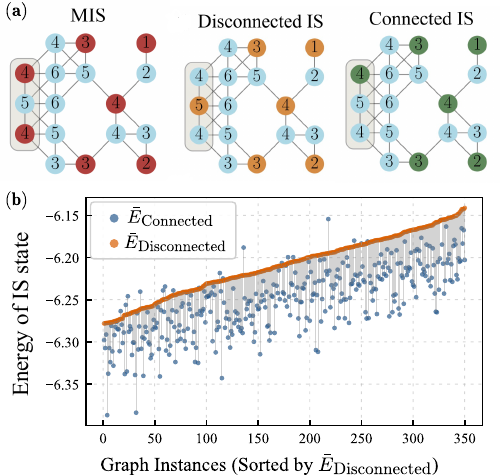}
    \caption{\textbf{(a)} A representative graph marking the MIS (red) and two independent sets (IS) of size $|\text{MIS}|-1$: a connected IS, which is subset of the MIS (green), and a disconnected IS, which is not (yellow). The shaded nodes indicate vertices that are not common to all three sets.
    \textbf{(b)}  Energies of  atomic configurations with $|\text{MIS}| -1$ excitations representing connected (blue) and disconnected (orange) ISs. Data collected for 1000 random graphs and sorted by the average energy of disconnected ISs. }
    \label{fig:figure4}
\end{figure}

To introduce our modified definition of the HP, we first divide the collection of  independent sets of size $\alpha-1$   into two: \emph{connected sets} --- which are subsets of a MIS, and \emph{disconnected sets} --- which are not. Representative examples are shown in \figref{figure4}(a). States representing disconnected sets act as ``traps'' in the adiabatic evolution that hinder reaching the MIS state. This is because they
 differ from MIS states by excitations on three or more atoms rather than a single atom in connected IS states.
While the energies of all atomic configurations with $\alpha-1$ excitations are degenerate in the traditional global-drive protocol, the LD-AQC lifts the degeneracy.
  \figrefbegin{figure4}(b) shows that the energy of dis-connected ISs is higher on average than that of connected ISs. This is desirable, because by increasing the energy gap between MIS states and disconnected ISs, the population in such states is suppressed.  
  
  Next, we show that the HP essentially enumerates the fraction of disconnected sets out of all ISs of size $\alpha-1$. This understanding will assist us in generalizing the HP definition.  Let $\mathcal{I}_{\alpha-1}$ be the set of all independent sets of size $\alpha - 1$. The traditional HP can be rewritten as
\begin{equation}
    \mathcal{HP}_{\text{trad}} = \frac{\sum_{S \in \mathcal{I}_{\alpha-1}} 1}{\sum_{S \in \mathcal{I}_{\alpha-1}} c_S}.
    \label{eq:hardness_parameter_standard_equivalent} 
\end{equation}
The equivalence of this definition with \eqref{hardness_parameter_standard}  is proven in \appref{EM-HPproof}. The sums in the definition run over all independent sets $S \in \mathcal{I}_{\alpha-1}$. The parameter $c_S$ denotes the number of vertices that can be added to an  IS  of size $\alpha - 1$ to form a MIS; $c_S = 0$ for  disconnected sets, $c_S = 1$ for connected sets that are a subset of a single MIS, and $c_S > 1$ for connected sets that are subsets of multiple MISs. The generalized HP is defined as \begin{equation}
    \mathcal{HP}_{\text{LD}} \equiv \frac{\sum_{S \in \mathcal{I}_{\alpha-1}} w_S}{\sum_{S \in \mathcal{I}_{\alpha-1}} w_S \cdot c_S}, \quad w_S \equiv \frac{1}{|E_S - E_{\text{MIS}}|},
    \label{eq:hardness_parameter_LD}
\end{equation}
where the weights $w_S$ quantify the deviation of the state energy $E_S$ from the average MIS energy, $E_{\text{MIS}}$. All weights are finite following the energy condition [\eqref{energy-cond}]. 
With this definition, states whose energy is closer to the average  MIS energy dominate the sum and, hence, dictate the problem hardness.  
Alternative definitions are discussed in \appref{HP_anlyz}. 
The generalized hardness parameter, $\mathcal{HP}_{\mathrm{LD}}$ reduces to the traditional metric $\mathcal{HP}_{\mathrm{trad}}$ in the limit of homogeneous detuning (\appref{HP_reduc}).  The modified HP [\eqref{hardness_parameter_LD}]
can be used to analyze the time complexity of the LD-AQC for solving MIS but applies to a more general class of weighted-MIS problems (WMIS).

\section{Discussion and Outlook}\label{sec:discussion}

We presented an efficient Hamiltonian engineering algorithm for solving MIS. In contrast to previous work from \citeasnoun{ebadi2022quantum} that used \emph{global driving pulses} to perform the AQC, our method relies on \emph{local degree-dependent detuning profiles}. By fitting numerical data obtained from simulations of our algorithm on hundreds of random MIS instances, we find that the success probability metric---defined as the negative logarithm of the infidelity---exhibits a power-law dependence on the problem hardness [as defined in \eqref{hardness_parameter_LD}]. Furthermore, we demonstrate a clear scaling advantage of the local-drive method over the global-drive approach (\figref{figure3}). This advantage is expected, as local degree-dependent drives exploit graph information through a preprocessing step. The nontrivial point, however, is that this preprocessing remains efficient, scaling as $\mathcal{O}(N^2)$, i.e., polynomial in the number of vertices. We attribute the acceleration of our method to the energetic penalization of trap states during the annealing schedule (\figref{figure4}). We note that this advantage fundamentally relies on the variance in the local degree. For $d$-regular graphs the detuning factors become spatially uniform, and our protocol naturally reduces to homogeneous driving.

To conclude, we mention directions for future investigation. The performance is expected to improve by optimizing the local detuning profiles, e.g., via quantum optimal control~\cite{koch2022quantum}. The framework presented here also naturally extends to other combinatorial optimization problems, such as MAX-cut~\cite{farhi2012performance}. While our simulations explored 11 and 12-atom systems, future work should investigate larger systems, either by using tensor network simulations~\cite{verstraete2004renormalization,orus2014practical} or by using real quantum hardware. Finally, although demonstrated within the context of neutral-atom arrays, our approach is applicable, more generally, to other qubit platforms.

\section*{Acknowledgment}
AP acknowledges support from the Israel Science Foundation Grant No. 1484/24, from the Center for Theory of Quantum Computing funded by VATAT, and the GIF Nexus Grant No. I-1581-303.7/2026. The authors thank Daniel Turyansky for insightful discussions regarding numerical simulations of AQC with Rydberg arrays.

\bibliographystyle{quantum}
\bibliography{LDmis}

\clearpage
\onecolumn
\appendix


\section{Proof of propositions in Algorithm 1}\label{app:proof-alg1}

Algorithm 1 (\algref{Algo1}) guarantees that energy manifolds of $k$ excitations are gapped for all $k$ by finding $k^*$ and $\beta^*$ such that the energy difference function [\eqref{energy-cond}] is positive, i.e., $D_{k^*}(\beta^*)>0$. The proof relies on two statements: First, that the function $D_k(\beta)$ has a global minimum with respect to $k$, i.e., there exists a $k^*$ for which $D_k(\beta)\geq D_{k^*}(\beta)\,\forall \,k$ and, second, that if there exists a critical $\beta_{c}$ for which the minimum vanishes, then by choosing $\beta^*<\beta_{c}$ we get $D_k(\beta^*)>0\,\forall\,k$. To prove these statements, we introduce the difference of energy differences
\begin{equation}
    \varepsilon_k(\beta) \equiv D_{k+1}(\beta) - D_k(\beta).
    \label{eq:diff-Da}
\end{equation} 
The function $\varepsilon_k(\beta)$ is the discrete gradient of $D_k(\beta)$, which is evident from the writing: $\varepsilon_k\equiv\frac{D_{k+1} - D_k}{\Delta k}$ with $\Delta k = 1$. Next, we introduce claim:

\begin{proposition}[Discrete Gradient Identity]\label{prop:prop1}
The following identity holds:
\begin{equation}
\varepsilon_k(\beta) = f_{N-k-1}(\beta) - f_{k-1}(\beta).
\label{eq:simplified-varepsi}
\end{equation}
\end{proposition}

\textbf{Proof:} The proof follows from substituting the definition of $D_k(\beta)$ [\eqref{D_k_definition}] into the definition of $\varepsilon_k(\beta)$ [\eqref{diff-Da}] and the cancellation of terms:
\begin{align}
D_{k+1}(\beta)-D_k(\beta)
&= \left( \sum_{i=N-(k+1)}^{N-1} f_i - \sum_{i=0}^{(k+1)-2} f_i \right)
- \left( \sum_{i=N-k}^{N-1} f_i - \sum_{i=0}^{k-2} f_i \right) \nonumber \\
&= \left( f_{N-k-1} + \sum_{i=N-k}^{N-1} f_i - \sum_{i=N-k}^{N-1} f_i \right)
- \left( f_{k-1} + \sum_{i=0}^{k-2} f_i - \sum_{i=0}^{k-2} f_i \right) \nonumber \\
&= f_{N-k-1}(\beta) - f_{k-1}(\beta).
\label{eq:energy-diff-equality}
\end{align}
This completes the proof. $\blacksquare$

Since $f_{k}$ is monotonic, several properties of $\varepsilon_k$ follow:

\begin{proposition}[Monotonicity and Zero-Crossing]\label{prop:prop2}
\begin{enumerate}
    \item $\varepsilon_k$ is monotonically increasing, that is $\varepsilon_{k+1}\ge \varepsilon_k\,\forall\,k$.
    \item $\varepsilon_k$ changes sign from $\varepsilon_1\le 0$ to $\varepsilon_{\lfloor N/2\rfloor+1}\geq 0$.
    \item There exists a critical $k^*\leq{\lfloor \tfrac{N}{2}\rfloor+1}$ such that $\varepsilon_{k^*}\geq 0$.
\end{enumerate}
\end{proposition}

\textbf{Proof:} Since $f_{k-1} \geq f_k$ and $f_{N-k-1}\leq f_{N-k-2}$, it follows that $f_{k-1} - f_{N-k-1}\geq f_k-f_{N-k-2}$, which implies that $\varepsilon_{k+1}\geq\varepsilon_k$ (from \eqref{diff-Da}). Moreover, $\varepsilon_1 = D_2 - D_1 = f_{N-2} - f_0\leq 0$ and $\varepsilon_{\lfloor N/2\rfloor+1} = f_{N-\lfloor N/2\rfloor - 1} - f_{\lfloor N/2\rfloor-1}\geq 0$ from monotonicity of $f(\beta,d_{i})$. $\blacksquare$

\begin{corollary}[Global Minimum of $D_k(\beta)$]\label{cor:global_min}
Since $\varepsilon_k$, representing the slope of $D_k$, is monotonically increasing and crosses 0 at $k^*$, it follows that $D_k(\beta)$ has a global minimum at $k^*$, i.e.,
\begin{equation}
D_{k^*}(\beta)\leq D_k(\beta) \quad \forall\, k,\beta.
\end{equation}
\end{corollary}

While our algorithm seeks $\beta_{c}$ which satisfies $D_{k^*}(\beta_{c}) = 0$, we do not prove its existence for a general function $f(\beta,d_{i})$. We merely state that if it exists, then \eqref{energy-cond} holds. A condition for its existence is detailed in \appref{min_exist}.

\section{Alternative definition of \texorpdfstring{$\mathcal{HP}_{\text{trad}}$}{HP\_trad} [\texorpdfstring{\eqref{hardness_parameter_standard}}{Eq. 7}]}\label{app:EM-HPproof}

In this section, we prove the equivalence of the two expressions for $\mathcal{HP}_\text{trad}$ in \eqref{hardness_parameter_standard}:
\begin{equation}
    \frac{D_{|\text{MIS}|-1}}{|\text{MIS}| \cdot D_{|\text{MIS}|}} = \frac{\sum_{j\in{|\text{MIS}|-1}} 1}{ \sum_{j\in{|\text{MIS}|-1}} c_j}.
\end{equation}
First, consider the numerators. The degeneracy $D_{|\text{MIS}|-1}$, which represents the total number of valid ISs of size $|\text{MIS}|-1$, is written on the right-hand side (RHS) by enumerating all states in that set, $\sum_{j\in{|\text{MIS}|-1}} 1$. As for the denominators, we show below:
\begin{equation}
|\text{MIS}| \cdot D_{|\text{MIS}|} = \sum_{j\in{|\text{MIS}|-1}} c_j.
\label{eq:denominators}
\end{equation}
To this end, we describe two ways for counting the number of ways an IS of size $|\mathrm{MIS}|-1$ can be extended to a MIS by flipping a single vertex. This is essentially the number of connected ISs --- defined as subsets of a MIS of size $|\text{MIS}|-1$ --- where ISs that are subsets of multiple MISs are counted with multiplicity. Let us call this number \emph{the IS-to-MIS extension count}.

To prove the left-hand side, we adopt a ``top-down'' perspective: For each set in the MIS group, by removing any vertex out of its $|\text{MIS}|$ vertices, one obtains a connected IS of size $|\mathrm{MIS}|-1$. Consequently, the total number of such sets is $|\text{MIS}| \cdot D_{|\text{MIS}|}$, where $D_{|\text{MIS}|}$ denotes the degeneracy of the MIS manifold.

To prove the right-hand side, one may enumerate these sets using a ``bottom-up'' approach: For each IS of size $|\text{MIS}|-1$ denoted by index $j$, let $c_j=0$ if it is disconnected and $c_j\geq 1$ if it is connected and belongs to $c_j$ MISs. The IS-to-MIS extension count is thus given by $\sum_{j \in \{|\text{MIS}|-1\}} c_j$, which completes our proof. $\blacksquare$

\section{Testing different detuning profiles}\label{app:EM-DiffDetuning}

\begin{figure}[htbp]
    \centering
    \includegraphics[width=0.85\textwidth]{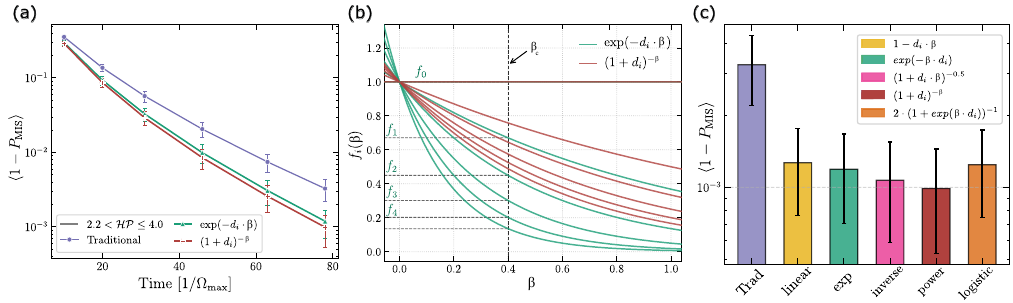}
    \caption{\textbf{(a)} Average infidelity $\langle 1 - P_{\text{MIS}} \rangle$ versus total protocol time for specific hardest graphs subset ($2.2 < \mathcal{HP} \le 4.0$). The traditional homogeneous AQC (purple) is compared against local detuning utilizing exponential (green) and inverse-power (red) functions.
    \textbf{(b)} Dependence of the local detuning function $f(\beta,d_{i})$ on the control parameter $\beta$. The green curves correspond to the exponential profile $f(\beta,d_{i}) = \exp(-d_i \beta)$, while red represent the power-law profile $f(\beta,d_{i}) = (1+d_i)^{-\beta}$, evaluated for $d_i = 0, \dots, 5$.
    \textbf{(c)} Average infidelity $\langle 1 - P_{\text{MIS}} \rangle$ for $t_f \cdot \Omega_{\text{max}} = 25\pi$ for a graph ensemble, comparing six scaling functions $f(\beta,d_{i})$.}
    \label{fig:figure_EM2}\label{fig:figure5}
\end{figure}

To investigate the sensitivity of the algorithmic performance to the specific functional form of $f(\beta,d_{i})$, we compare several candidate profiles—ranging from linear to exponential and power-law dependencies—while enforcing three consistency requirements: (i) the functions must satisfy the conditions:
\begin{subequations}
\label{eq:f_conditions}
\begin{gather}
d_0\leq d_1\leq\dots\leq d_{N-1}{\label{eq:deg-array}}\\
f(\beta,d_0)\geq f(\beta,d_1)\geq\dots\geq f(\beta,d_{N-1})>0,
\end{gather}
\end{subequations}
(ii) they are normalized such that $f(\beta,d_{i}) = 1$ for the homogeneous case ($d_i = 0$); and (iii) the functions remain positive, $f(\beta,d_{i}) \in [0,1]$, throughout the interval $\beta \in [0,1]$. The results of this functional comparison are presented in \figref{figure_EM2}.

Panel (a) illustrates the average infidelity $\langle 1 - P_{\text{MIS}} \rangle$ as a function of the protocol time $t_f$ for a sub-ensemble of hardness instances ($2.2 < \mathcal{HP} \le 4.0$). While all local addressing schemes provide a significant speedup over the traditional AQC baseline (purple), the power-law profile $f(\beta,d_{i}) = (1+d_i)^{-\beta}$ (red) on average demonstrates the most favorable performance. The specific profiles $f(\beta,d_{i})$ for varying $d_i$ are depicted in \figref{figure_EM2}(b). Finally, we compare six parameterization functions $f(\beta,d_{i})$ at a fixed protocol time $t_f \cdot \Omega_{\text{max}} = 25\pi$. The comparison in \figref{figure_EM2}(c) confirms that the LD approach is robust for the chosen functions. Every tested profile achieves a significant reduction in the failure probability, with the power-law profile emerging as the optimal protocol in this set of functions.

\section{Statistics of the performance}\label{app:EM-PerformStas}

\begin{figure}[htbp]
    \centering
    \includegraphics[width=0.7\textwidth]{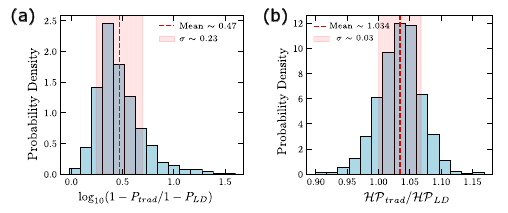}
    \caption{Statistical analysis of the data presented in \figref{figure3}. \textbf{(a)} Probability density of the logarithmic error ratio, $\log_{10} [ (1 - P_{\text{trad}}) / (1 - P_{\text{LD}}) ]$.
    \textbf{(b)} Probability density of the hardness parameter ratio, $\mathcal{HP}_{\text{trad}} / \mathcal{HP}_{\text{LD}}$.}
    \label{fig:figure_EM1}
\end{figure}

To evaluate the performance of the LD-MIS algorithm for different MIS instances, we show in this section the distribution of results obtained for the ensemble of graphs tested in \figref{figure3} in the text. Specifically, we examine in \figref{figure_EM1}(a) the distribution of the logarithmic error ratio, defined as $\log_{10} [ (1 - P_{\text{trad}}) / (1 - P_{\text{LD}}) ]$, to compare the failure probabilities of the traditional and LD protocols. We find that this ratio is positive across all instances, with a mean value of $0.47$, showing that the LD protocol outperforms the traditional approach on all instances. A mean value of $0.47$ implies a nearly triple reduction in the average error rate.

In \figref{figure_EM1}(b), we quantify the reduction in algorithmic complexity by analyzing the ratio of the hardness parameters, $\mathcal{HP}_{\text{trad}} / \mathcal{HP}_{\text{LD}}$, across the same ensemble of graph instances. We find that our LD protocol typically yields a smaller hardness parameter than the traditional homogeneous AQC, with an average reduction in effective problem hardness of $\sim 3.5\%$.

\section{Existence Condition for a Root of \texorpdfstring{$D_k(\beta)$}{D\_k(a)}}\label{app:min_exist}\label{SM:sec:proofs:min_exist}

We claim in the text that if a parameter $\beta_{c}$ exists such that $D_{k^*}(\beta_{c}) = 0$, then by choosing $\beta^*<\beta_{c}$ we get $D_k(\beta^*)>0\,\forall\,k$ and the energy condition is met [\eqref{energy-cond}]. Here, we show a condition for its existence. Let the sequence $f(\beta,d_{i})$ be non-increasing and positive. For $k \in \{0, \dots, N\}$, the function $D_k(\beta)$ satisfies the following decomposition:
\begin{equation}
    D_k(\beta) = f(\beta,d_{N-k}) + \sum_{i=0}^{k-2}\left(f(\beta,d_{N-k+1+i}) - f(\beta,d_{i})\right).
\end{equation}
Since $f(\beta,d_{N-k+1+i}) \le f(\beta,d_{i})$, every term in the sum is non-positive, yielding the upper bound $D_k(\beta) \le f_{N-k}(\beta)$. Strict positivity holds if and only if $f(\beta,d_{N-k}) > \sum_{i=0}^{k-2}\big(f(\beta,d_{i})-f(\beta,d_{N-k+1+i})\big)$. Hence, the latter inequality needs to be violated in order to find a parameter $\beta^*$ such that $D_k(\beta^*) = 0$.

\section{A condition on the detuning sweep range and the blockade interaction}\label{app:blockade_check}\label{SM:sec:proofs:blockade_check}

While attempting to preferentially excite atoms with small degrees, we must ensure that our protocol does not excite neighboring atoms, i.e., that the blockade interaction always dominates the energy and suppress neighboring excitations. This condition is formulated in this section.

In principle, the blockade energy -- determined by the shortest distance between atoms in the array, $U_0$ -- must be larger than the maximum possible energy reward for exciting a single atom. In other words the weighted detunings must not overcome the physical Rydberg blockade. We must ensure that two adjacent atoms $i, j$ are never excited.

The energy of an invalid state containing excitations of neighboring atoms $i$ and $j$ is $E_{ij} = -\frac{\delta_0}{2}(f_i+f_j) + U$. The energy of a valid state with atom $i$ excited is $E_i = -\frac{\delta_0}{2} f_i$. Requiring that $E_{ij} > E_{i}$, we obtain:
\begin{equation}
    -\frac{\delta_0}{2}(f_i+f_j) + U > -\frac{\delta_0}{2} f_i \quad\Rightarrow\quad U > \frac{\delta_0}{2} f_j.
\end{equation}
To be safe for all pairs, we require $U > \frac{\delta_0}{2} f_i$ for all $i$. This gives the conservative bound:
\begin{equation}
    U > \frac{\delta_0}{2} \, f_{\max}.
\end{equation}

\section{The reduction of \texorpdfstring{$\mathcal{HP}_\text{LD}$}{HP\_LD} to \texorpdfstring{$\mathcal{HP}_\text{trad}$}{HP\_trad} in limit of homogeneous detuning}\label{app:HP_reduc}\label{SM:sec:proofs:HP_reduc}

In this section we show that the generalized definition for the hardness parameter, $\mathcal{HP}_\text{LD}$, reduces to the traditional definition, $\mathcal{HP}_\text{trad}$, in the limit of homogeneous detuning.

In the homogeneous limit, the energy of IS states with a fixed number of excitations is degenerate. Hence, the difference between the IS and MIS energies, $|E_i - E_{\text{MIS}}|$, is constant for all states $i$ in the IS manifold. Therefore, the weight factor in the definition of $\mathcal{HP}_\text{LD}$ [$w_i$ in \eqref{hardness_parameter_LD}] is constant for all $i$-th ($w_i = w_c$). We insert this constant weight into the definition of $\mathcal{HP}_{\text{LD}}$:
\begin{align}
    \mathcal{HP}_{\text{LD}} &= \frac{\sum_{\text{|MIS|-1}} w_c}{\sum_{\text{|MIS|-1}} w_c \cdot c_{i}} \nonumber \\
    &= \frac{w_c \cdot \sum_{\text{|MIS|-1}} 1}{w_c\cdot \sum_{\text{|MIS|-1}} c_{i}} \nonumber \\
    &= \frac{w_c \cdot (\text{Number of MIS-1 states})}{w_c \cdot (\text{Number of connected MIS-1 states})} \nonumber \\
    &= \frac{w_c \cdot D_{|\mathrm{MIS}|-1}}{w_c \cdot|\mathrm{MIS}| \cdot D_{|\mathrm{MIS}|}} \nonumber \\
    &= \frac{ D_{|\mathrm{MIS}|-1}}{|\mathrm{MIS}| \cdot D_{|\mathrm{MIS}|}} = \mathcal{HP}_{\text{trad}}.
\end{align}
This demonstrates that our energy-weighted $\mathcal{HP}_{\text{LD}}$ generalizes the original degeneracy-based $\mathcal{HP}_{\text{trad}}$, reducing to it exactly when the non-uniformity is removed.

\section{Relating success probability metric to the traditional \texorpdfstring{$\mathcal{HP}$}{HP}}\label{app:SPM_HP}\label{SM:sec:proofs:SPM_HP}

In this section we establish the scaling of the success probability metric of the LD-AQC protocol with the traditional hardness parameter, following \figref{figure3} in the main text. We observed that the LD-AQC performance scales with the traditional protocol according to a power law:
\begin{equation}
\mathrm{SPM}_{\mathrm{LD}} \propto (\mathrm{SPM}_{\mathrm{trad}})^{0.75}.
\end{equation}
Substituting the fit for $\mathrm{SPM}_{\mathrm{trad}}$ from \figref{figure3}(c):
\begin{equation}
\mathrm{SPM}_{\mathrm{LD}} = a^{0.75}_{\text{trad}} \mathcal{HP}_{\text{trad}}^{-0.75b_{\text{trad}}}. 
\end{equation}
Using the definition of the SPM:
\begin{equation}
\mathrm{SPM}_{\mathrm{LD}} = -\ln(1 - P_{\text{MIS}}^{\text{LD}}),
\end{equation}
we thus obtain
\begin{equation}
\ln(1 - P_{\text{MIS}}^{\text{LD}}) = -a^{0.75}_{\text{trad}} \mathcal{HP}_{\text{trad}}^{-0.75b_{\text{trad}}}.
\end{equation}
To solve for $P_{\text{MIS}}^{\text{LD}}$, we exponentiate both sides of the equation obtaining
\begin{equation}
e^{\ln(1 - P_{\text{MIS}}^{\text{LD}})} = e^{- a^{0.75} _{\text{trad}}\mathcal{HP}_{\text{trad}}^{-0.75b_{\text{trad}}}}.
\end{equation}
Finally, we obtain the scaling relation used in the main text:
\begin{equation}
P_{\text{MIS}}^{\text{LD}} = 1 - \exp\left( -a^{0.75}_{\text{trad}} \mathcal{HP}_{\text{trad}}^{-0.75b_{\text{trad}}} \right).
\end{equation}

\section{Alternative definitions of the generalized hardness parameter}\label{app:HP_anlyz}\label{SM:sec:HP_anlyz}

\subsection{Definition of the weighting factor \texorpdfstring{$c_j$}{c\_j}}\label{app:HP_anlyz_multy_vs_binary}

A central question in the development of the generalized hardness parameter ($\mathcal{HP}$) was whether connected states (those in the $|\text{MIS}|-1$ manifold that can transition to an MIS configuration) should be weighted by their multiplicity. Specifically, rewriting below the definition for $\mathcal{HP}_{\text{LD}}$ from the main text:
\begin{equation}
    \mathcal{HP}_{\text{LD}} \equiv \frac{\sum_{j\in\text{|MIS|-1}} w_j}{\sum_{j\in\text{|MIS|-1}} w_j \cdot c_{j}},
\label{eq:HP_multiplicity}
\end{equation}
we explored two alternative definitions for $c_j$: (1) where $c_j = 0,1,\dots$ represents the number of distinct MIS configurations accessible from state $j$, and (2) where $\tilde{c}_j \in \{0, 1\}$, indicates only whether a state is connected to at least one MIS configuration or not, oblivious to its multiplicity.

To identify the most suitable definition for the $\mathcal{HP}$, we calculated the correlation between each $\mathcal{HP}$ candidate and the minimal spectral gap $\Delta_{\text{min}}$, assuming that the definition with the highest correlation should be chosen. We used four correlation metrics:
\begin{enumerate}
    \item Spearman: $r_s = \frac{\mathrm{cov}[\mathrm{R}[X], \mathrm{R}[Y]]}{\sigma_{\mathrm{R}[X]} \sigma_{\mathrm{R}[Y]}}$,
    \item Pearson: $\rho_{X,Z} = \frac{\mathrm{cov}(X, Z)}{\sigma_X \sigma_Z}$, with $Z=\log(\Delta_{\mathrm{min}})$,
    \item Distance correlation: $\mathrm{dCor}(X,Y) = \frac{\mathrm{dCov}(X,Y)}{\sqrt{\mathrm{dVar}(X)\mathrm{dVar}(Y)}}$, and
    \item Mutual information: $I(X;Y) = \sum_{x \in \mathcal{X}} \sum_{y \in \mathcal{Y}} p(x,y) \log\left(\frac{p(x,y)}{p(x)p(y)}\right)$.
\end{enumerate}

We compared three $\mathcal{HP}$ candidates:
\begin{enumerate}
    \item $\mathcal{HP}_1$ (the LD metric with multiplicity count, $c_j = 0,1,\dots$),
    \item $\mathcal{HP}_2$ (the LD metric with a binary indicator: $c_j = 0, 1$), and
    \item $\mathcal{HP}_3$ (the traditional metric, with $w_j = 1, c_j = 0,1,\dots$).
\end{enumerate}

As summarized in \tableref{correlations}, the local detuning metric incorporating multiplicity ($\mathcal{HP}_1$) consistently exhibits the strongest correlation across all statistical measures.

\emph{This result suggests that not all connected states contribute equally to the adiabatic transition probability; rather, configurations that interface with a larger number of MIS states play a more critical role in the manifold's dynamics.} Consequently, the multiplicity of these connections is a necessary component for accurately quantifying the effective algorithmic complexity.

\begin{table}[htbp]
\centering
\caption{Correlation metrics between different $\mathcal{HP}$ definitions and the minimal spectral gap $\Delta_{\min}$.}
\label{tab:correlations}
\begin{tabular}{lccc}
\toprule
Correlation metric & $\mathcal{HP}_1$ ($c_j = 0,1,\dots$) & $\mathcal{HP}_2$ ($c_j = 0,1$) & $\mathcal{HP}_3$ (Traditional) \\
\midrule
Spearman ($\rho$)      & \textbf{--0.9410} & --0.9121 & --0.8785 \\
Pearson ($\log$)       & \textbf{--0.9223} & --0.8922 & --0.8609 \\
Distance correlation   & \textbf{0.9118}  & 0.8764  & 0.8458  \\
Mutual information (nats)     & \textbf{1.2293}  & 0.9378  & 0.7942  \\
\bottomrule
\end{tabular}
\end{table}

The significant superiority of $\mathcal{HP}_1$ is most evident in the mutual information (MI) score, which measures the total statistical dependence (including non-linear effects) between the variables. The fact that $\mathcal{HP}_1$ yields an MI of $1.23$ compared to $0.9378$ for the binary metric indicates that the multiplicity of connections to the MIS manifold is a primary factor in determining the complexity.

\subsection{\texorpdfstring{$\mathcal{HP}$}{HP} and Minimal Gap Analysis}\label{app:HP_anlyz_HP_vs_mingap}\label{SM:sec:HP_anlyz:HP_vs_mingap}

\begin{figure}[htbp]
    \centering
    \includegraphics[width=0.85\textwidth]{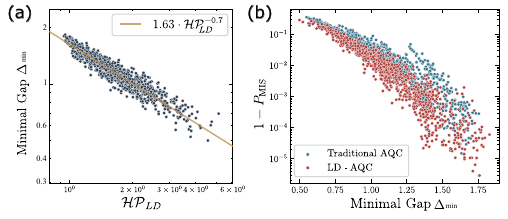}
    \caption{(a) The minimal spectral gap $\Delta_\text{min}$ evaluated as a function of the local detuning hardness parameter $\mathcal{HP}_{\mathrm{LD}}$. The fit ($\sim \mathcal{HP}_{\mathrm{LD}}^{-0.7}$) shows the negative correlation.
    (b) The infidelity $1 - P_{\text{MIS}}$ mapped against the minimal energy gap for individual graph instances. While both traditional AQC and the LD-control protocol show increased infidelity for smaller gaps, the LD approach systematically lowers the error for similar minimal gap regions.}
    \label{fig:figure_EM3}\label{fig:min_gap}\label{SM-fig:figure_EM3}
\end{figure}

The performance of adiabatic protocols is fundamentally dictated by the minimal spectral gap, $\Delta_{\text{min}}$, encountered during the evolution. To validate $\mathcal{HP}_{\mathrm{LD}}$ as a good indicator of algorithmic difficulty, we analyze its correlation with the spectral properties of the system Hamiltonian. As illustrated in \figref{figure_EM3}(a), the minimal spectral gap exhibits a strong inverse correlation with our hardness parameter. The numerical data across the graph ensemble follows a power-law scaling of $\Delta_{\text{min}} \propto \mathcal{HP}_{\mathrm{LD}}^{-0.7}$, confirming that instances with higher $\mathcal{HP}_{\mathrm{LD}}$ values are characterized by smaller minimal gap. This relationship justifies the use of $\mathcal{HP}_{\mathrm{LD}}$ as an indicator of the minimal gap.

Furthermore, we examine the relationship between the final infidelity, $1 - P_{\text{MIS}}$, and the minimal gap in \figref{figure_EM3}(b). While both the traditional and local detuning (LD) protocols show the expected trend of increasing error as the gap closes, the LD-AQC protocol systematically achieves lower infidelity compared to the traditional approach for the same values of $\Delta_{\text{min}}$.

\end{document}